\documentclass[aps,pra, twocolumn,showpacs,reprint,superscriptaddress]{revtex4-1}

\usepackage{graphicx}
\usepackage{hyperref}
\usepackage{amsmath}
\usepackage{amssymb}

\begin{document}

\title{High-efficiency WSi superconducting nanowire single-photon detectors\\ for quantum state engineering in the near infrared}

\author{Hanna Le Jeannic}
\thanks{These authors contributed equally.}
\affiliation{Laboratoire Kastler Brossel, UPMC-Sorbonne Universit\'es, CNRS, ENS-PSL Research University, Coll\`ege de France, 4 Place
Jussieu, 75005 Paris, France}

\author{Varun B. Verma}
\thanks{These authors contributed equally.}
\affiliation{National Institute of Standards and Technology, 325 Broadway, Boulder, CO 80305, USA}

\author{Adrien Cavaill\`es}
\affiliation{Laboratoire Kastler Brossel, UPMC-Sorbonne Universit\'es, CNRS, ENS-PSL Research University, Coll\`ege de France, 4 Place
Jussieu, 75005 Paris, France}
\author{Francesco Marsili}
\affiliation{Jet Propulsion Laboratory, California Institute of Technology,
4800 Oak Grove Dr., Pasadena, California 91109, USA}
\author{Matthew D. Shaw}
\affiliation{Jet Propulsion Laboratory, California Institute of Technology,
4800 Oak Grove Dr., Pasadena, California 91109, USA}
\author{Kun Huang}
\affiliation{Laboratoire Kastler Brossel, UPMC-Sorbonne Universit\'es, CNRS, ENS-PSL Research University, Coll\`ege de France, 4 Place
Jussieu, 75005 Paris, France}
\author{Olivier Morin}
\affiliation{Max-Planck-Institut f\"ur Quantenoptik, Hans-Kopfermann-Str. 1, D-85748 Garching, Germany}
\author{Sae Woo Nam}
\affiliation{National Institute of Standards and Technology, 325 Broadway, Boulder, CO 80305, USA}

\author{Julien Laurat}
\email{julien.laurat@upmc.fr}
\affiliation{Laboratoire Kastler Brossel, UPMC-Sorbonne Universit\'es, CNRS, ENS-PSL Research University, Coll\`ege de France, 4 Place
Jussieu, 75005 Paris, France}

\begin{abstract}
We report on high-efficiency superconducting nanowire single-photon detectors based on amorphous WSi and optimized at 1064 nm. At an operating temperature of 1.8 K, we demonstrated a 93\% system detection efficiency at this wavelength with a dark noise of a few counts per second. Combined with cavity-enhanced spontaneous parametric down-conversion, this fiber-coupled detector enabled us to generate narrowband single photons with a heralding efficiency greater than 90\% and a high spectral brightness of $0.6\times10^4$ photons/(s$\cdot$mW$\cdot$MHz). Beyond single-photon generation at large rate, such high-efficiency detectors open the path to efficient multiple-photon heralding and complex quantum state engineering.
\end{abstract}

\maketitle

The near-infrared wavelength range has been an important playground for a large community in quantum optics, in particular at $\lambda=1064$~nm. At this wavelength that corresponds to Nd:YAG lasers, narrow-linewidth and ultra-stable lasers are available, and also ultralow-loss optical coatings and photodiodes with close-to-unity efficiency. Pioneering \cite{Wu} and world-record \cite{Eberle2010} demonstrations of squeezed light \cite{Andersen2016} have been performed at this wavelength, as well as quantum teleportation \cite{Bowen2002PRA} and enhanced quantum metrology, including for gravitational wave detection \cite{Ligo2013}. In contrast to these realizations, the preparation of single-photon states or more complex states involving larger photon-number components generally requires single-photon detections to herald their preparations \cite{Illuminati,Morin2014,Huang2015}. The efficient detection of single photons is therefore a fundamental and demanding requisite for these investigations \cite{Hadfield2009}. This is particularly true for multiple heraldings, a challenging trend for developing novel quantum state engineering capabilities and further scalability in quantum networks \cite{Walmsley}.

However, detecting efficiently single photons at $\lambda=$~1064~nm has been a long-standing issue. Avalanche photodiodes (APDs) based on Silicon are limited to a few percent quantum efficiency. InGaAs/InP ones, even though large improvements have been obtained in the recent years with these devices, still exhibit limited efficiency for this specific wavelength and often suffer from large dark count rates if used in free-running mode. In this wavelength range, the first generation of NbN superconducting nanowire single-photon detectors \cite{Hadfield} (SNSPDs) was typically limited to 20-30$\%$ efficiencies. 

Recently, the development of SNSPDs based on tungsten silicide (WSi) \cite{Baek2011, Verma2012} enabled to outperform other infrared single-photon detectors, despite great progresses with NbN-based devices \cite{Rosenberg2013,Miki2013}. In addition to a better internal efficiency than other materials, the amorphous nature of WSi facilitates the fabrication of large sensitive area and the embedding of the material inside an optical stack to enhance absorption. System detection efficiency (SDE) greater than 90\% has been achieved in the wavelength range $\lambda=1520-1610$ nm \cite{Marsili}. Efficient operation above 2 K has also been demonstrated at 1310 nm \cite{Verma2014}. 

In this paper, we report on the optimization of WSi SNSPD at $\lambda=$ 1064 nm. A system detection efficiency of $\eta_{det}=93\pm3\%$ is achieved with a dark noise below a few counts per second. As an application, we show the fast heralding of narrowband single photons. Based on a 53-MHz bandwidth optical parametric oscillator (OPO) with a close-to-unity escape efficiency, a heralding efficiency $P_1$ (i.e. the probability of finding a single photon in a given spatio-temporal mode per heralding event) greater than 90\% and a heralding rate up to 1 MHz are demonstrated. The spectral brightness, $\sim0.6\times10^4$ photons/(s$\cdot$mW$\cdot$MHz), has been improved by more than 15 fold relative to our previous realizations. This result makes our system one of the brightest parametric-down conversion sources to date \cite{Luo2015}, due to an intrinsic narrow bandwidth combined with a high detection efficiency for the heralding path. 
  
  \begin{figure}[t!]
\includegraphics[width=0.98\columnwidth]{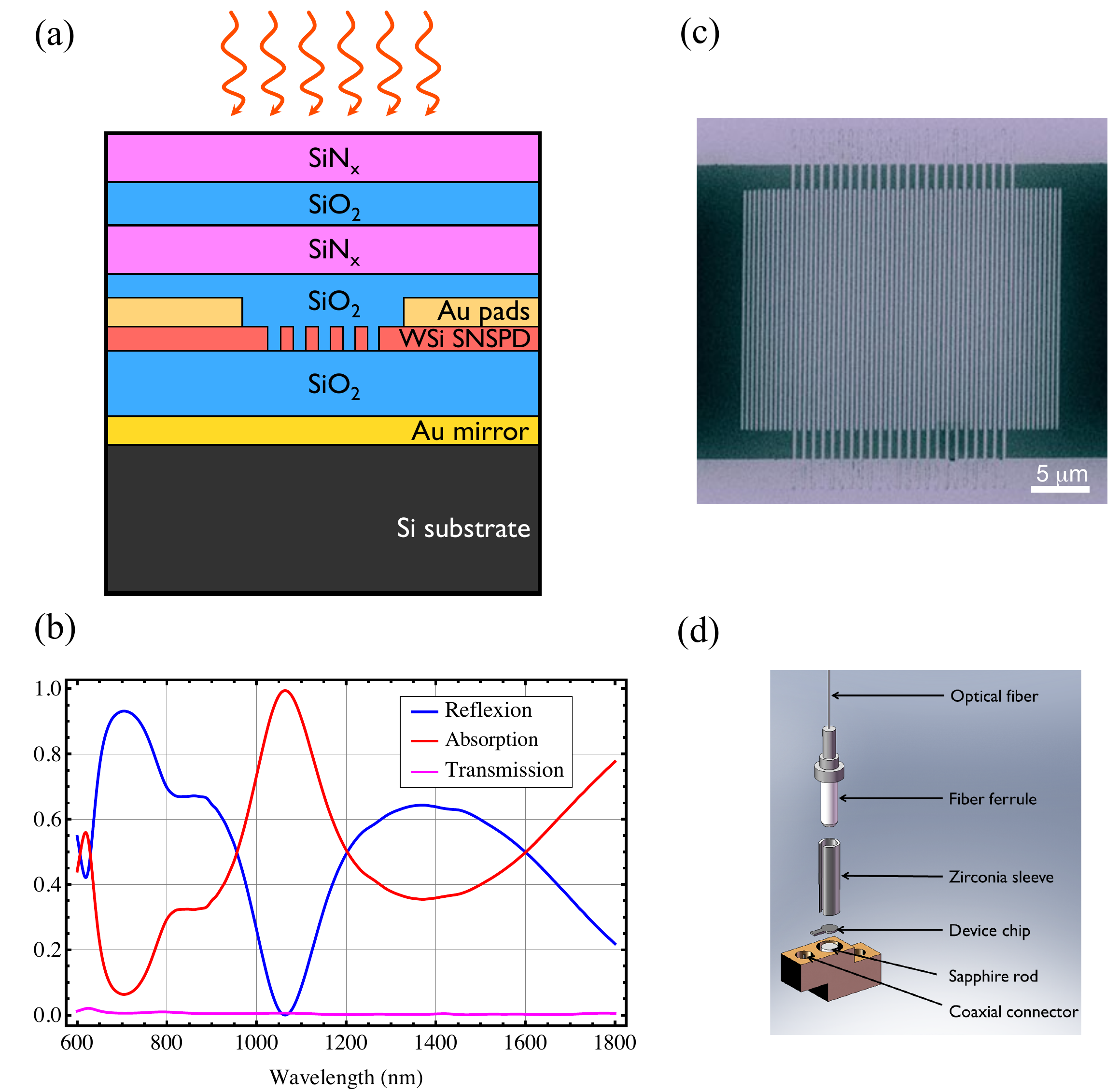}
\caption{(a) The WSi meander is embedded inside an optical stack deposited on a Si substrate. The layers have been optimized to enhance absorption at 1064 nm. An antireflection coating is deposited on the top surface and a quarter-wave spacer layer is realized between the meander and the bottom gold mirror. (b) Theoretical absorption, reflection and transmission spectra for the optimized layers. (c) SEM image of the meander. The nanowire has a width of 140 nm and pitch of 245 nm. (d) Detector assembly and fiber connection. The single-mode fiber is self-aligned via a zirconia sleeve. The 16 $\mu$m x 16 $\mu$m active area of the SNSPD is larger than the 10 $\mu$m mode field diameter of a standard single-mode fiber, to allow for slight misalignment.}
\label{fig1}
\end{figure} 

The W$_{0.8}$Si$_{0.2}$ detectors, as shown in Fig. \ref{fig1}, were optimized for maximum absorption at a wavelength of 1064 nm. Simulations were performed via a Matlab program running RCWA (rigorously coupled-wave analysis) to optimize the layer thicknesses, given the optical constants of the materials (Fig. \ref{fig1}b). The first step of the fabrication process consists in the deposition of a 80 nm-thick gold mirror on a 3'' Si wafer by electron beam evaporation and lift-off, with a 2 nm Ti adhesion layer below and above the mirror. A quarter-wave spacer layer consisting of 152 nm of SiO2 was then deposited by plasma-enhanced chemical vapor deposition (PECVD), and Ti/Au contact pads were patterned by electron beam evaporation and lift-off. The $\sim4$ nm-thick WSi superconductor layer was deposited by DC magnetron cosputtering from separate W and Si targets at room temperature, and capped with 2 nm of amorphous Si to prevent oxidation. Photolithography and etching in an SF6 plasma were used to define a 20 $\mu$m-wide strip between gold contact pads. Electron beam lithography using PMMA resist and etching in SF6 were then used to define nanowire meanders with a width of 140 nm and pitch of 245 nm within the 20 $\mu$m-wide strip. An SEM image of the meander is shown on Fig. \ref{fig1}c. An antireflection coating was deposited on the top surface consisting of 102 nm SiO2, 137 nm SiNx, 171 nm SiO2, and 192 nm SiNx. Finally, a keyhole shape was etched through the Si wafer around each SNSPD, which could then be removed from the wafer and self-aligned to a single-mode optical fiber to within $\pm$ 3 $\mu$m \cite{Miller}. The keyhole part is attached on the top of a sapphire rod and bonded using aluminium microwires to a piece of copper in contact with an RF cable, as shown in Fig. \ref{fig1}d.

The SNSPD is mounted at the bottom of a double-wall dipstick, which is immersed in a liquid Helium dewar and pumped to achieve a 1.8 K operation temperature. A 1 meter-long single-mode optical fiber is connected to the detector and embedded inside the dipstick. The fiber end face is AR coated. Voltage pulses resulting from photon detection are first filtered by a 100 MHz low-pass filter then amplified via a chain of two 1 GHz amplifiers with 24 dB of gain each (Minicircuits  ZFL-1000LN+), all operating at room temperature. 

To quantify the overall system detection efficiency $\eta_{det}$, which includes fiber and connexion losses, we use a strongly attenuated continuous-wave Nd:YAG laser with a mean photon-number of about 100,000 photons/s. As the detector is polarization sensitive (typically up to 10\% difference in efficiency), the polarization of the probe light is optimized to maximize the number of counts. The main uncertainty is coming from the power-meter accuracy (Ophir PD300-IRG), i.e. 3\%. The  standard error on the count rate and the laser power fluctuations were measured to be below the percent level. 

\begin{figure}[t]
\centering
\includegraphics[width=0.95\columnwidth]{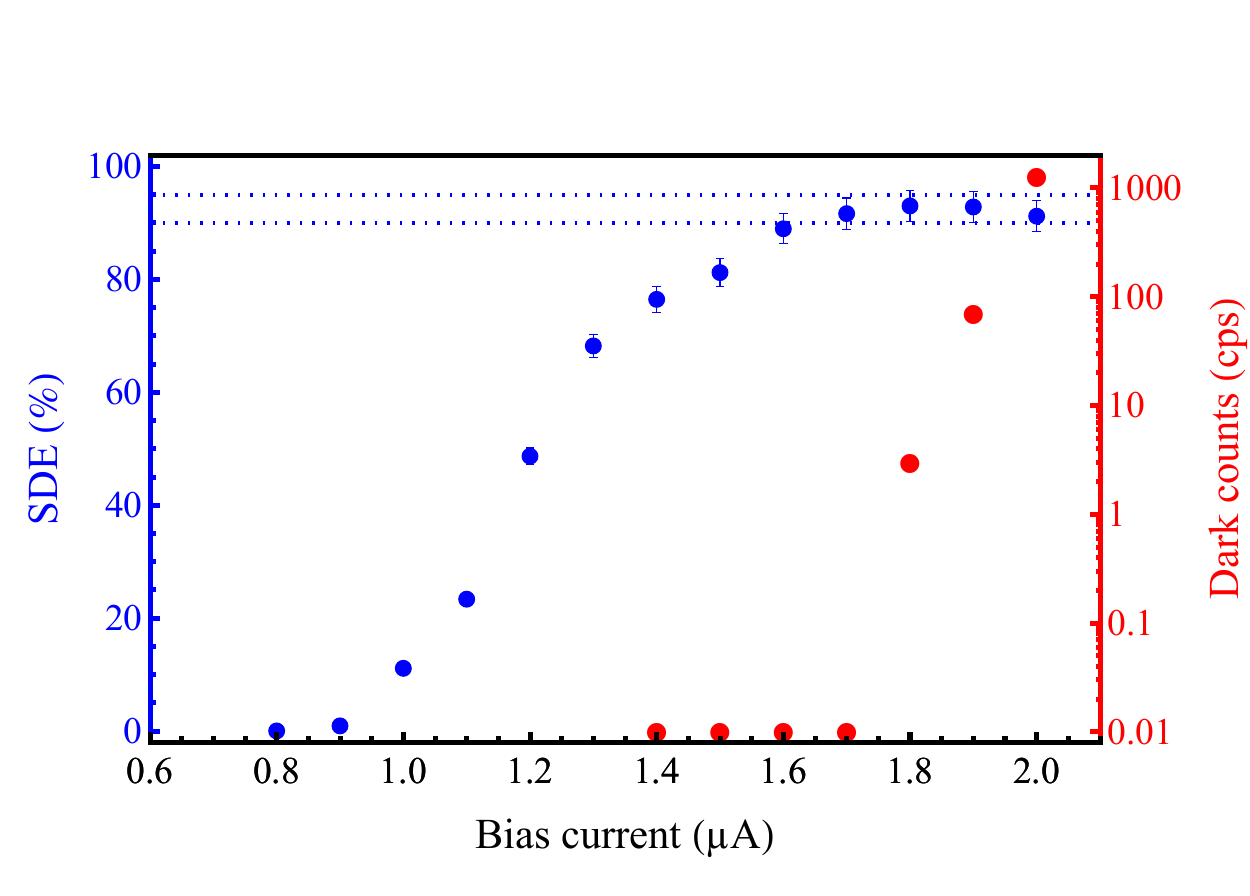}
\caption{System detection efficiency (SDE) at 1064 nm (blue) and dark counts (red) as a function of the bias current. For a bias current of $1.8\, \mu$A, a value equal to 0.9 times the switching current, a SDE of $\eta_{det}=93~\pm~3\%$ is obtained with a dark noise of 3 cps. These measurements are obtained at an operating temperature of 1.8~K.}
\label{fig:figure2}
\end{figure}

Figure \ref{fig:figure2} provides the measurement of the SDE and dark count rate as a function of the bias current. As it can be seen, at this operating temperature the system reaches the inflection current leading to a saturated efficiency before reaching the switching current, $\sim 2\,\mu$A, for which the device switches to the normal non-superconducting state. At a bias current of $1.8\,\mu$A, an SDE of $\eta_{det}=93\pm3\%$ is obtained with a dark noise limited to 3 counts per second (cps). For bias currents closer to the switching current, the dark count rate increases rapidly.  

We now turn to an application, namely the use of this optimized detector for heralding the generation of single photons. Preparing high-purity single photons at a large rate and into a well-controlled spatio-temporal mode is a crucial requirement for a variety of quantum information protocol implementations \cite{Eisaman2011,Gisin}. We use here a general conditional preparation method based on twin beams \cite{Mandel,Lvovsky2001,DAuria2012} where the detection of one photon of the pair heralds the presence of the second one \cite{Nielsen2007,Mosley2008,Benson2009,Wolgramm2011,Ramelow2012,Pomarico2012,Krapick2013,Fekete2013,Fortsch2013,Sasaki2015,Furusawa2015,Ngah2015,Luo2015}.

\begin{figure}[t!]
\centering
\includegraphics[width=0.95\columnwidth]{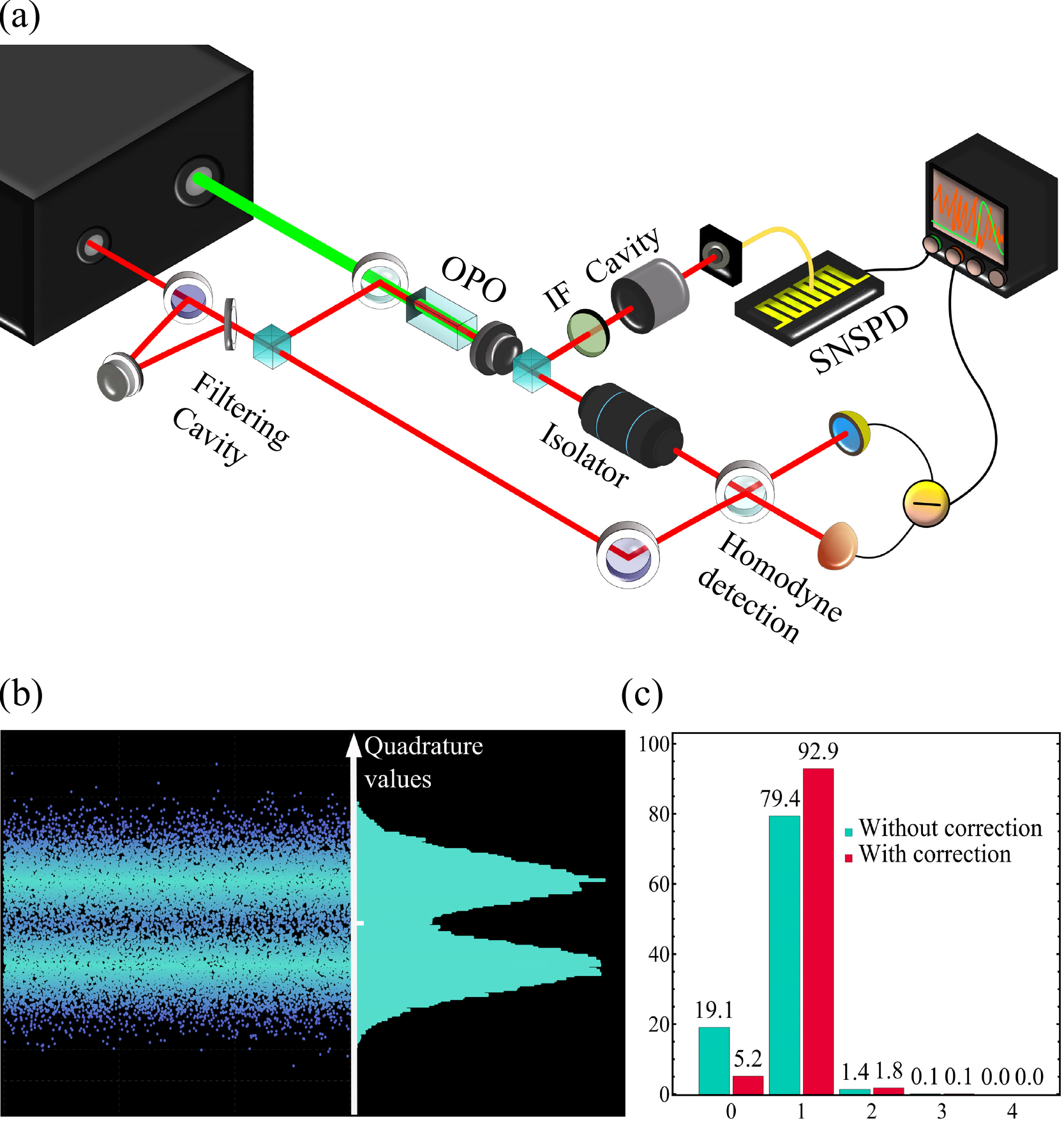}
\caption{(a) Experimental setup for the heralded single-photon source. A polarization non-degenerate and triply-resonant OPO is pumped far below threshold by a continuous-wave laser at 532 nm. Due to the cavity enhancement, the pump power is in the mW range. At the output of the OPO, photons pairs are separated via a polarizing beam splitter. Single-photon detection on the conditioning path heralds the emission of its twin, which is then characterized by homodyne detection. The overall transmission of the conditioning path, which includes frequency filtering (interferential filter IF and resonant cavity), reaches 50\%. An optical isolator enables to avoid any backscattering from the detection system. (b) Quadrature values and distribution measured via the homodyne detection. (c) Photon-number distribution for a 1mW pump power, with and without correction from detection losses (15\%). The corrected heralding efficiency, i.e. the probability of obtaining a single photon per heralding event at the output of the OPO, reaches $P_1=$93\%. The vacuum admixture is limited to 5\%.}
\label{fig:figure3}
\end{figure}

The experimental setup is sketched in Figure \ref{fig:figure3}a and has been described elsewhere \cite{Morin2012,Jove,Morin2013a}. It is based on a triply-resonant optical parametric oscillator made of a semi-monolithic linear cavity with on one side a 1cm-long coated KTP crystal ($R=95\%$ for 532 nm and HR for 1064 nm) and on the other side a curved output mirror (HR for 532 nm and $R=90\%$ for 1064 nm). This design provides a close-to-unity escape efficiency $\eta_{OPO}= T/(T+L)=0.96$ with $T$ the transmission of the output coupler and $L$ the residual intracavity losses. This parameter quantifies the intrinsic loss of the source and a value close to unity enables to obtain a very low admixture of vacuum, and therefore a large heralding efficiency. In complex state engineering, this low admixture enables to generate high-purity states with a large negativity of the associated Wigner function \cite{Morin2014}.

In our scheme, photon pairs emitted at 1064 nm with orthogonal polarizations are separated on a polarizing beam splitter. The idler photons are then frequency filtered by an interferential filter ($0.5$ nm bandwidth) and a Fabry-Perot cavity (300 MHz bandwidth), to select photons only in the central frequency mode of the OPO cavity. The overall transmission of the filtering path is measured to be $\sim 50 \%$. These photons are finally detected by the WSi SNSPD and herald the generation of single photons on the signal mode. The heralded photons are then characterized using balanced homodyne detection. A few-mW laser beam (local oscillator, filtered by a high-finesse cavity) is mixed with the heralded state. Interference visibility above 99\% is achieved. The two beam-splitter outputs are detected via two photodiodes (97\% quantum efficiency) and the photocurrents are then subtracted. This difference gives access to the quantum fluctuations of the signal that can be used to reconstruct the density matrix via a maximum-likelihood algorithm. The electronic noise is 20 dB below the shot noise level at the central frequency (4\% equivalent loss) and the propagation loss between the OPO and the detection is measured to be $\simeq 6\%$. The overall detection loss amounts therefore to 15\%. Figure \ref{fig:figure3}b shows the measured quadrature values and the associated distribution for a 1mW pump power. The temporal mode of the heralded state used here is a double-decaying exponential profile, as determined by the 53-MHz bandwidth of the OPO cavity \cite{Morin2013b}.

\begin{figure}[t!]
\centering
\includegraphics[width=0.95\columnwidth]{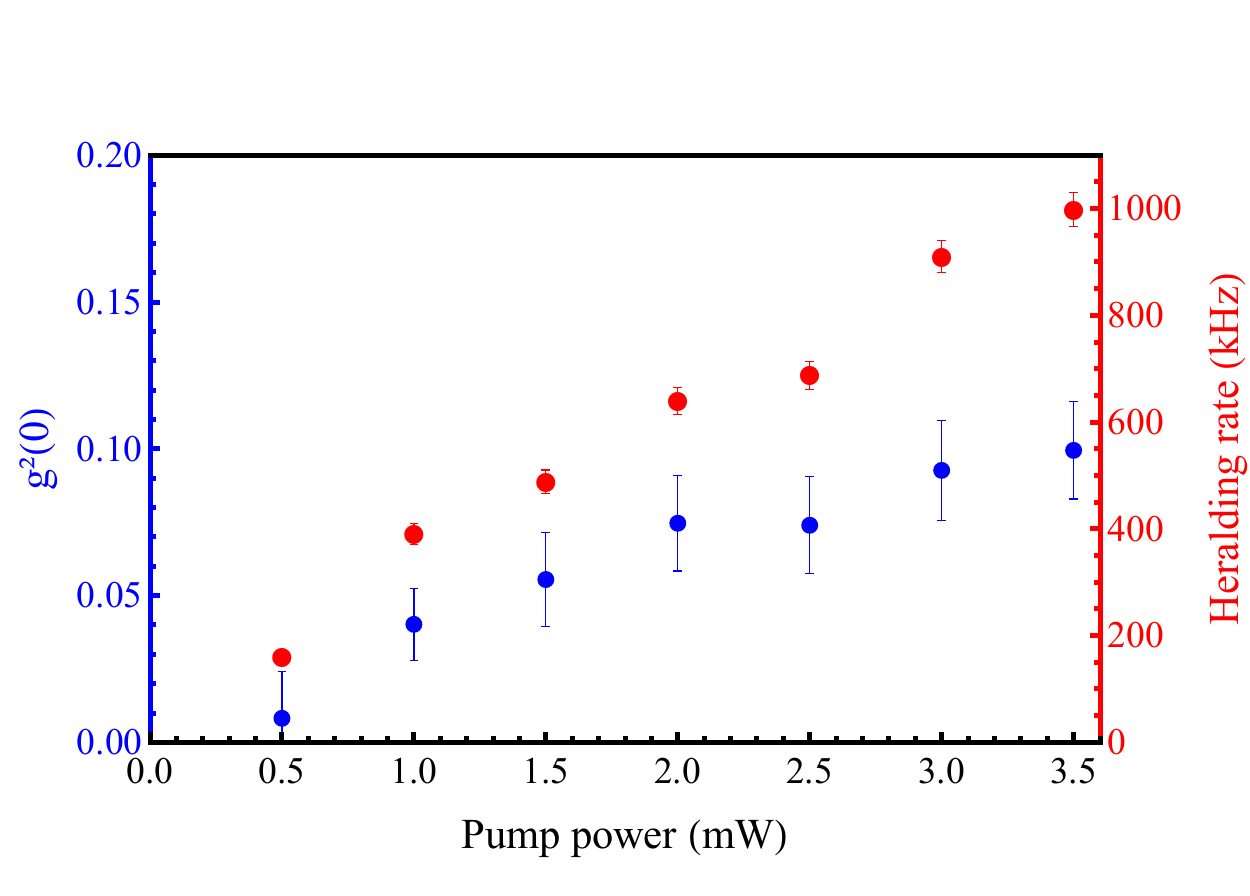}
\caption{Heralding rate and conditional autocorrelation function $g^{(2)}(0)$ as a function of the pump power. A spectral brightness as large as $0.6\times10^4$ photons/(s$\cdot$mW$\cdot$MHz) is obtained, close to the maximal achievable value in this parametric down-conversion system.}
\label{fig:figure4}
\end{figure}

Figure \ref{fig:figure3}c provides the reconstructed diagonal elements of the density matrix, i.e. here the weight of the different photon-number components, including the vacuum component. The single-photon component reaches $79\pm 0.5\%$ without any correction from detection losses.  By taking into account detection losses (i.e. $15\%$), the single-photon weight is estimated to be $P_1=93 \pm 0.5\%$. This value corresponds to the corrected heralding efficiency, i.e. the probability of finding a single photon in a given spatio-temporal mode at the OPO output per heralding event. The residual vacuum component is as low as $5\pm 0.5\%$, a value consistent with the OPO escape efficiency.

Figure \ref{fig:figure4} finally provides the heralding rate and the conditional autocorrelation function $g^{(2)}(0)$ for the heralded state as a function of the continuous-wave pump power. The correlation function can be calculated from the measured photon-number components as $g^{(2)}(0)=\frac{\langle(a^{\dagger})^2a^2\rangle}{{\langle a^{\dagger}a\rangle}^2}\simeq\frac{2\,P_2}{{(P_1+2P_2)}^2}$ with $P_1$ and  $P_2$ the single- and two-photon components. This parameter, which goes to zero for single-photon Fock state, increases with the unwanted multi-photon components. As a typical parametric down-conversion scheme, rate and $g^{(2)}(0)$ have to be compromised. By increasing the pump power, we varied the heralding rate from 150 kHz to 1 MHz, with a $g^{(2)}(0)$ ranging from 0.008 to 0.1. 

The achieved spectral brightness is equal to $0.6\times10^4$~photons/(s$\cdot$mW$\cdot$MHz). This value corresponds to a $\sim 15$-fold improvement relative to our previous works \cite{Morin2012}. Notably, this value is close to the maximal achievable value in this parametric down-conversion system, namely $\sim1.2\times10^4$ photons/(s$\cdot$mW$\cdot$MHz) if one assumes perfect single-photon detector and no loss in the heralding path (50\% overall loss). 

In conclusion, we reported the optimization of high-efficiency WSi superconducting nanowire single-photon detector at 1064~nm. At 1.8 K, the system detection efficiency reaches 93\% with a background noise limited to a few counts per second. This efficiency is the highest demonstrated to date for this wavelength. Using this SNSPD to herald the generation of single photons, we demonstrated a bright single-photon source that gathers a heralding efficiency larger than 90\% and a narrowband spectrum. The unique combination of large escape efficiency OPOs, which enable a very low admixture of vacuum, and superconducting single-photon detectors with close-to-unity efficiency, as achieved here, will make protocols based on multiple-photon conditioning more accessible and scalable. It also opens the path to a variety of realizations in the developing optical hybrid approach to quantum information where discrete- and continuous-variables states, operations, and toolboxes are combined \cite{Andersen2015}. Our recent demonstration of high-rate generation of large squeezed optical Schr\"odinger cat states that relies on double heralding is a first example \cite{Huang2015}.

\begin{acknowledgments}
This work was supported by the European Research Council (Starting Grant HybridNet). Part of this research was carried out at the Jet Propulsion Laboratory, California Institute of Technology, under a contract with the National Aeronautics and Space Administration. V.B.V. and S.W.N. acknowledge partial funding for detector development from the DARPA Information in a Photon (InPho) and QUINESS programs. J.L. is a member of the Institut Universitaire de France.
\end{acknowledgments}

\bibliography{sample}
{}
\end{document}